# A NOVEL STATISTICAL DIAGNOSIS OF CLINICAL DATA

Gene Kim and MyungHo Kim

**Abstract.** In this paper, we present a diagnosis method of diseases from clinical data. The data are routine test such as urine test, hematology, chemistries etc. Though those tests have been done for people who check in medical institutes, how each item of the data interacts each other and which combination of them cause a disease are neither understood nor studied well. Here we attack the practically important problem by putting the data into mathematical setup and applying support vector machine. Finally we present simulation results for fatty liver, gastritis etc and discuss about their implications.

## §1. Introduction

Since the human genome sequence was completed, there has been a lot of excitement in the hopes of deciphering the sequences and discovering new drugs for diseases. However, the obtained results did not meet the expectations because researchers were not successful in developing a data analysis method that is suitable for the current situation, and there is no standard method to analyze the great amount of genome data. As a result, scientists have not been able to take full advantage of the complete human genome sequence.

So the new concepts and novel approach for analyzing genetic data such as SNP, genetic markers and DNA chip data etc. are needed. More precisely, there is a need to develop a new method and a concept that deals with many variables simultaneously, instead of dealing with a variable individually.

Along this line, we introduce a new concept in the emerging area of bioinformatics (See [2]) and apply it to clinical data for appropriate diagnosis and analysis, combining with machine-learning methods.

This new approach could open up a new horizon to medical diagnosis and enhance health care for persons. The idea and goal are straightforward. Traditionally, doctors set a normal range of blood pressure based on data obtained from a large number of people. If a patient is excluded from the range, the doctors tried to adjust it to the "normal range". Over the years, people have observed the fact that some healthy people are not in the "normal range." This fact implies that there are other factors than blood pressure that "cooperate" with the blood pressure factor to keep a person's health in balance. This makes us develop a new concept of analyzing multiple variables (contributing factors) simultaneously, not individually.

We start with two concepts:
1. In order to classify objects we are interested in, we need to find a new way of representing the objects into numbers.
2. To get a criterion (cutoff) used to divide a group into subgroups, a knowledge–based method (machine learning methods such as support vector machine, neural network, decision tree etc.) is needed.

Following the concepts above, we represent a group of objects as vectors. Then we label them and separate the group into two subgroups. From the division, we obtain a cutoff/criterion distinguishing one subgroup from the other subgroup, and the cutoff will be used to determine, to which subgroup, a new vector representation of an object belongs to.

In §2, we explain a way of representation into a vector and in §3, we will review the support vector machine known as the most powerful classification method. Finally we will discuss about the practical simulation and perspective in §4.

## §2. Representation as Vectors

Here, we will present a way of representation of clinical data into numbers, i.e. vectors and clinical tests mean tests being done in medical institute, for example, a blood test, urine test or MRI etc.

### §2.1 Procedure of representation

Suppose we have $M$ clinical test results $C_1, C_2... C_M$. Then we can represent the results as a vector as described below:.

Step 1: For each $C_i$,

Case 1: $C_i$ is a number

    Take it as a component of the vector;

Case 2: $C_i$ represents one of certain *k* stage

    Take one of any chosen *k* distinct numbers as a component of the vector. For example, for the WBC test in urine having 5 stages such as Positive 1, Positive 2, Positive 3, Negative, Trace, we may choose 5 distinct numbers, for example, 1, 2, 3, 4, 5 and take each of them as a component of the clinical test vector, representing the stages Positive 1, Positive 2, Positive 3, Negative, Trace respectively.

Step 2: Enumerate the numbers in the step 1 and consider it as the clinical test vector in the *M* dimensional Euclidean space.

Note that the methods described above can be applied for any clinical data in diagnosing a disease or some states, which will be discussed below.

### §2.2 Labeling vectors

Once we have numericalized the clinical data of persons (or organisms), we label each vector +1 or –1, accordingly. More precisely, suppose we have a group of persons (or organisms) and represent them as vectors. We can label the vectors depending on various cases. The characters "I" and "II" will refer to some groups, which will vary depending on the context.

Here are a few examples of labeling vectors accordingly.

(1)     Depending on whether the person (or the organism) has a specific disease or not, the vector is labeled by +1 or –1 respectively.

(2)     Given a disease, depending on whether the disease status of persons (or organisms) is at the stage, "I" or "II", the vector is labeled by +1 or –1 respectively.

(3)     It is believed that each person has his/her own degree of radiation sensitivity due to genetic difference that may be distinguished by clinical data. Label a vector +1, if the person has the degree of radiation sensitivity, "I", and

otherwise –1.

(4) Some people have some allergies against a medicine while some do not. Label a vector +1 if the person has an adverse effect and otherwise –1.

To make things simple and clear for explanation, we will restrict ourselves to the case (1). Thus, after we have represented persons or organisms as vectors in the $M$ dimensional Euclidean space, each of those vectors is labeled +1 or –1, wherein the labeled vector +1 indicates a disease and the labeled vector –1 indicates absence of the disease.

By applying classification methods such as support vector machine, neural network etc, we can find a cutoff (criterion) to separate the set of +1 labeled vectors from the set of –1 labeled vectors with optimal errors. More precisely, the cutoff is determined by a hypersurface dividing the Euclidean space into two disjointed sets and will be used for determining whether an unlabeled vector representing a person (or an organism) belong to one of those two sets, and accordingly the person will be diagnosed to have a disease or not. It works similarly for the cases (2), (3), and (4) above.

Suppose a cutoff hypersurface separates a Euclidean space into two sets, "I" and "II". Also, suppose that "I" set contains more +1 labeled vectors than "II", while "II" set do more –1 labeled vectors than "I". We mean optimal errors by maximizing the percentage of the set of +1 labeled vectors in "I" among the total number of labeled vectors of "I" and the percentage of the set of –1 labeled vectors in "II" among the total number of labeled vectors of "II". This is the optimal classification that we are referring to in the discussion below in §3, as well (Also refer to [4] for some details.).

Figs 1 – 4 in the Appendix show examples of a hypersurface separating labeled vectors in the 3-dimensional Euclidean space.

Finding such a hypersurface (for example, n- dimensional hyperplane, n-dimensional sphere etc. in the (n+1) dimensional Euclidean space) is a crucial step for classifying people into patient group and normal group. Furthermore, if necessary, in the classifying step, by repeating the use of machine learning methods to already divided set to get two subsets, we may have two sets of the Euclidean space each of which set consists of several pieces. In other words, the group classified as normal or the other need not be a connected set mathematically. See FIG. 3 and Fig 4 showing such examples.

A hyperplane, which is a specific type of a cutoff surface, may be calculated by using an optimization problem comprising the following, wherein each $y_i$ is +1 or −1 and $x_i$ is a vector:

Maximize:

$$W(\alpha) = \frac{1}{2} \sum_{i,j=1}^{l} y_i y_j \alpha_i \alpha_j (x_i \bullet x_j) - \sum_{i=1}^{l} \alpha_i$$

Under the conditions $\sum_{i=1}^{l} \alpha_i y_i = 0,$ and

$0 \leq \alpha_i \leq C, i = 1, 2 ... l,$ wherein C is a given constant

For the derivation of the quadratic function $W$, refer to [1] and [4].

## §3. Brief Review of Support Vector Machine

Let $R^n$ be the *n*-dimensional Euclidean space and let $A$ and $B$ be two sets of finite number of points in $R^n$. Our question is whether there is a systematic way of dividing $R^n$ into two disjoint sets so that $A$ and $B$ are contained in either of two sets.

In mathematical terms, is there a way of obtaining a function $f: R^n \rightarrow R$ such that

$$f(x) = 1; x \in A$$
$$= -1; x \in B$$

The simplest function we may think that fits the above description is a function of degree 1, i.e., a linear function. Let us assume that $A$ and $B$ are separable by such a function, which is of form $w \bullet x + b = 0;$ where $x$ is a variable and $w$ and $b$ are constant vectors in $R^n$. (Here • denotes the standard inner product in Euclidean space.) However since there exists such hyperplanes separating $A$ and $B$ infinitely many, for practical implementation, we need to choose a specific one from those infinitely many candidates, which is a hard problem. In the late 1970, Vapnik solved this problem elegantly by introducing the definition of the optimal hyperplane and making a bridge to a nonlinear programming. Vapnik's solution separates the sets *A* and *B* so that the distance between the closest vectors of the two sets to the plane is "maximal". This maximal condition determines a unique plane. (For more details, see [4]). The essential implementation to get the optimal plane is to form a nonlinear programming problem that is obtained by imposing appropriate restrictions, and applying the Kuhn-Tucker's necessary conditions. This nonlinear programming problem gives us a unique plane with its unique expression. Let us explain about the unique expression briefly to give some idea about practical implementation into computer programming.

Suppose we need to express a decimal number in term of a fraction. We know that it could be expressed in infinitely many ways of fractions, i.e., 100/200, 28/56... 4/8. However, under the constraint that numerator and denominator be relatively prime, then there is a unique fractional expression, namely, 1/2. In the same principle, though there are infinitely many planes for separation, in the respect of separating a data set into exactly same two groups (even for a single plane, it has infinitely many ways of expressions, for example, $x + y = 1$ and $2x + 2y = 2$ represent the same line.), the optimization problem deduced from an observation manages to get rid of this ambiguity.

More precisely, it starts with the elementary distance formula from a point to a plane and, for simplicity, let $n = 2$. Then the distance from a point $(x_0, y_0)$ to the line $ax + by + c = 0$ is given by

$$\frac{|ax_0 + by_0 + c|}{\sqrt{a^2 + b^2}}$$

Thus, if we have two points $(0, 2)$ and $(0, -2)$ separated, we need to solve the following minimization problem. In other words,

Minimize:

$$\sqrt{a^2 + b^2}$$

Under the constraints $\varepsilon_1(a0 + b2 + c) \geq 1$

$$\varepsilon_2(a0 + b(-2) + c) \geq 1,$$

where $\varepsilon_1 \varepsilon_2 = 1$ and $\varepsilon_i = 1$ or $-1$.

Observe that, if the condition $|ax_0 + by_0 + c| \geq 1$ is imposed, the distance increases, as $\sqrt{a^2 + b^2}$ decreases. This plays a role corresponding to finding the fraction of which numerator and denominator are relatively prime, while $\varepsilon_1 \varepsilon_2 = 1$ of the two constraints does a role for separation with maximal margin from both of points. Therefore the line $2y = 0$ will be the optimal hyperplane, intuitively, the line passing through the middle point, the origin.

In general, let $(x_1, y_1) \ldots (x_l, y_l)$ be a set of labeled vectors, where each $y_i$ denotes where each vector $x_i$ belongs and takes either $+1$ or $\_1$. The formulation from the observation can be stated as follows

$$\text{Minimize: } f(w) = \frac{1}{2}\|w\|^2$$

under the constraints, $y_i[(x_i \bullet w) + b] \geq 1$, $i = 1, 2 \ldots l$

One way of solving this optimization problem is to use the associated Lagrangian of which definition is as follows:

**Definition 1** Given the following optimization problem

$$\text{Minimize } f(x), x = (x_1, x_{2\ldots} x_l),$$

under the constraints $g_i(x) \geq 0$, $i = 1, 2 \ldots m$

its associated Lagrangian is defined by

$$L(x, \lambda) = f(x) - \sum_{i=1}^{m} \lambda_i g_i(x)$$

where, $\lambda = (\lambda_1 \ldots \lambda_m)$, Lagrangian multipliers.

Kuhn and Tucker proved the minimization problem is equivalent to solving its associated Lagrangian functional (See [3]), i.e., finding a global saddle point of its associated Lagrangian functional. And in our case, the associated Lagrangian is given by

$$L(w, b, \alpha) = \frac{1}{2}\|w\|^2 - \sum_{i=1}^{m} \alpha_i \lambda_i ( y_i[(x_i \bullet w) + b] - 1)$$

At the global saddle point, $L$ should be minimized with respect with to $w$ and $b$ and maximized with respect to $\alpha_i \geq 0$. As a result, we have familiar necessary conditions of first order derivatives, called Kuhn-Tucker necessary conditions. Substitution those conditions in the Lagrangian functional leads us to the quadratic programming as mentioned in §2:

Maximize:

$$W(\alpha) = \frac{1}{2} \sum_{i,j=1}^{l} y_i y_j \alpha_i \alpha_j (x_i \bullet x_j) - \sum_{i=1}^{l} \alpha_i$$

Under the conditions $\sum_{i=1}^{l} \alpha_i y_i = 0,$ and

$0 \leq \alpha_i \leq C, i = 1, 2 ... l,$ wherein C is a given constant

In the case when the data set is not separable linearly, to construct a hyperplane of the optimal type, we introduce non-negative *slack* variables $\varepsilon_i$'s to constraints to reduce the sum of "distance of separation" errors.

Minimize: $f(w) = \frac{1}{2}\|w\|^2 + C \sum_{i=1}^{l} \varepsilon_i$

under the constraints, $y_i[(x_i \bullet w) + b] \geq 1 - \varepsilon_i, i = 1, 2 ... l$

Once again, the same arguments described above give the quadratic programming,

Maximize:

$$W(\alpha) = \frac{1}{2} \sum_{i,j=1}^{l} y_i y_j \alpha_i \alpha_j (x_i \bullet x_j) - \sum_{i=1}^{l} \alpha_i$$

Under the conditions $\sum_{i=1}^{l} \alpha_i y_i = 0$, and

$0 \leq \alpha_i \leq C, i = 1, 2 \ldots l,$ wherein C is a given constant

The practical implementation of this quadratic optimization program was discussed in some details. (See [1])

### §4. Simulation and Its Implication

### §4.1. Simulation

In this section, we present a few of simulation results with the following clinical items.

| Routine Check-Up | | |
|---|---|---|
| Age | Sex | Pulses |
| Belly fat rate | Body fat rate | Blood type |
| Height | Weight | Lung function |
| **Blood Pressure** | | |
| Systolic | | Diastolic |
| **Visual Acuity** | | |
| Left | | Right |
| **Chemistries** | | |
| Glucose | BUN | Creatinine |
| Na | K | Chloride |
| Calcium | Inorganic-phosphorus | Protein total |
| Albumin | Globulin | A/G ratio |
| Alk. Phosphatase | SGOT | SGPT |
| Total Bilirubin | Direct Bilirubin | LDH |
| Uric acid | Gamma-GT | |
| **Lipid Profile** | | |
| Cholesterol | Triglyceride | HDL-Cholesterol |
| LDL-Cholesterol | | |
| **Hepatitis** | | |
| HBs Ab (EIA) | HBs Ag (EIA) | HBs Ab (HA) |
| HBs Ag (HA) | HBs Ab (RIA) | HBs Ag (RIA) |
| HCV Ab (EIA) | | |
| **Urinalysis** | | |
| PH | Protein | Glucose |
| Ketone | Billirubin | Blood |
| Nitrite | Urobillinogen | WBC |
| **Thyroid** | | |
| Free T4 (RIA) | | TSH (RIA) |
| **Hematology** | | |
| WBC | RBC | Hb |
| HCT | MCV | MCH |
| MCHC | RDW | Platelet |
| Lymphocyte | Monocyte | Eosinophil |
| Basophil | Microscopy | |
| **Others** | | |
| Metamyelocyte | Myelocyte | Normblast |
| Promyelocyte | | |
| **STD Tests** | | |
| RPR | | |
| **Individual Tests** | | |
| B.M.D 1 | | PSA |

We started with the same number of normal and patients and obtained a single hyperplane to separate those two groups (See Fig. 5). What we want to see here is whether there is a conspicuous pattern for predicting a disease. In other words, whether there is a hyperplane drawing a line between normal and patient. Here are results.

1. Solidified breast test

a) We used mammography as our gold standard.
b) Patient means a person diagnosed as possible solidified breast with mammography test while normal means diagnosed as healthy with mammography test.
c) We choose 104 patients and 104 normal people and use the support vector machine.
d) As shown in the Table 1, the learning machine found out a hyperplane separating those two groups, i.e. patient and normal. One group, "Positive", contains 104 Patients and 0 Normals while the other group, "Negative", does 0 Patients and 104 Normals.

|          | Patient | Normal | Total |
| --- | --- | --- | --- |
| Positive | 104 | 0 | 104 |
| Negative | 0 | 104 | 104 |
| Total    | 104 | 104 | 208 |

Table 1

Sensitivity: 1.000000

Specificity: 1.000000

Predictive value for positive: 1.000000

Predictive value for negative: 1.000000

2. Fatty liver test

a) We used Ultrasound and U.G.I. as our gold standard.

b) Patient means a person diagnosed as possible fatty liver with those two tests while normal means diagnosed as healthy with the gold standard.

c) We choose 442 patients and 442 normal people and use the support vector machine.

d) As shown in the Table 2, the learning machine found out a hyperplane separating into two groups. One group, "Positive" contains 389 Patients and 37 Normals while the other group, "Negative" does 53 Patients and 442 Normals.

|          | Patient | Normal | Total |
|----------|---------|--------|-------|
| Positive | 389     | 37     | 426   |
| Negative | 53      | 405    | 458   |
| Total    | 442     | 442    | 884   |

Table 2

Sensitivity: 0.880090

Specificity: 0.916290

Predictive value for positive: 0.913146

Predictive value for negative: 0.884279

3. Gastritis test

a) We used endoscopy as our gold standard.

b) Patient means a person diagnosed as possible gastritis with the endoscopy test while normal means diagnosed as healthy with the gold standard.

c) We choose 592 patients and 592 normal people and use the support vector machine.

d) As shown in the Table 2, the learning machine found out a hyperplane separating into two groups. One group, "Positive" contains 451 Patients and 63 Normals while the other group, "Negative" does 141 Patients and 529 Normals.

|  | Patient | Normal | Total |
| --- | --- | --- | --- |
| Positive | 451 | 63 | 514 |
| Negative | 141 | 529 | 670 |
| Total | 592 | 592 | 1184 |

Table 3

Sensitivity: 0.761824

Specificity: 0.893581

Predictive value for positive: 0.877432

Predictive value for negative: 0.789552

What we have done here is to find a single optimal hyperplane to separate +1 labeled vectors from −1 labeled ones as shown in Fig. 5 below.

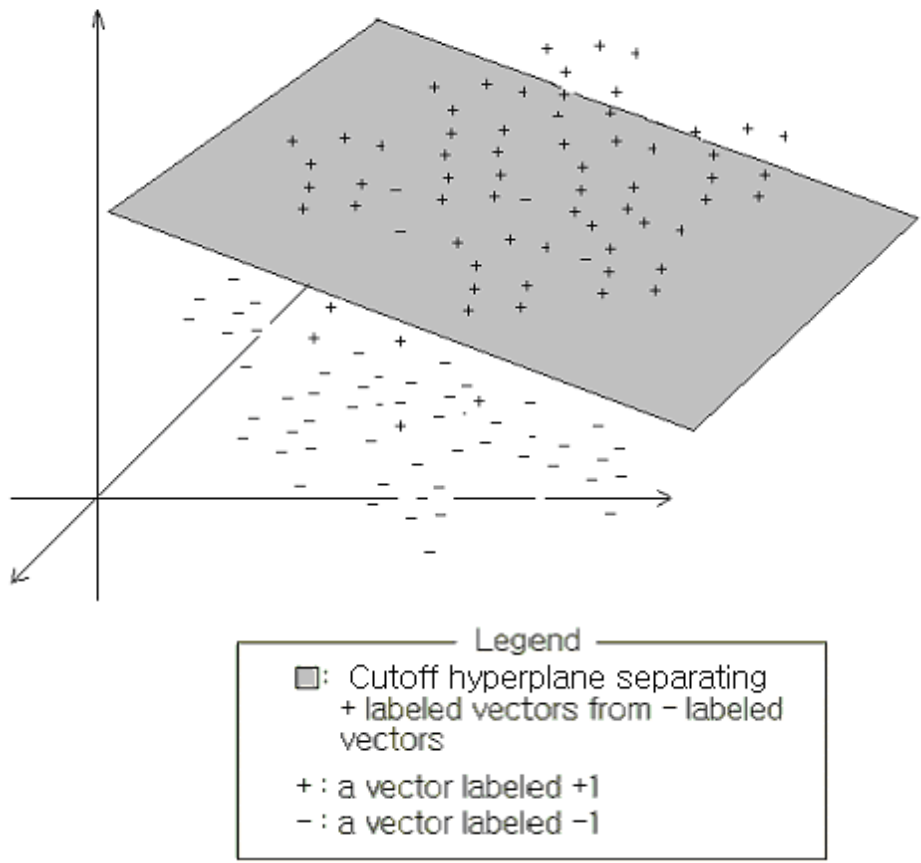

**Fig. 5**

## §4.2 Implication

From the routine check-up tests, we might be able to predict possible presence of several diseases simultaneously. As we listed in §4.1, there are many items that are being tested widely in the medical institutes. However, which combination of those familiar items is responsible for a disease is not understood well, for scientist have been searching for a single factor or element that is responsible for some disease or trait, which is not true for most cases. As clinical data pile up, the classification of the data will be necessary for each disease and its status, with respect to those test items. So in the near future, for each person who takes the routine check-up, a statistical diagnosis of diseases should be available for doctors, based on past clinical data. For example,

**Statistical Diagnosis with respect to the items above**

| Breast | 86% of healthy breast |
| --- | --- |
| Liver | 92.4 % of fatty liver |
| ... | ... |
| Gastritis | 74.5% of gastritis chance |

This diagnosis will be very helpful for doctors who have to care many patients within a limited time.

Note that the arguments applied to clinical data can be applied to any genetic data together with clinical data. (Refer to [2] for application of SNP data and refer to [5] for demonstration program.)

**References**

[1] Joachims, Thorsten, Making large-Scale SVM Learning Practical. Advances in Kernel Methods - Support Vector Learning, B. Scholkopf , C. Burges and A. Smola (ed.), MIT Press, 1999.

[2] Kim Gene and Kim MyungHo, Application of Support Vector Machine to detect an association between multiple SNP variations and a disease or trait, DIMACS workshop of Rutgers University, On the Integration of Diverse Biological Data, 2001. (http://dimacs.rutgers.edu/Workshops/Integration/program.html)

[3] Simmons, Donald, Nonlinear programming for operations research, Prentice Hall, 1975

[4] Vapnik, Vladimir. The Nature of Statistical Learning, Springer, New York, 1995

[5] http://www.biofront.biz/

Bioinformatics Frontier Inc. 93 B Taylor Ave, East Brunswick, NJ 08816
Email: mkim@biofront.biz

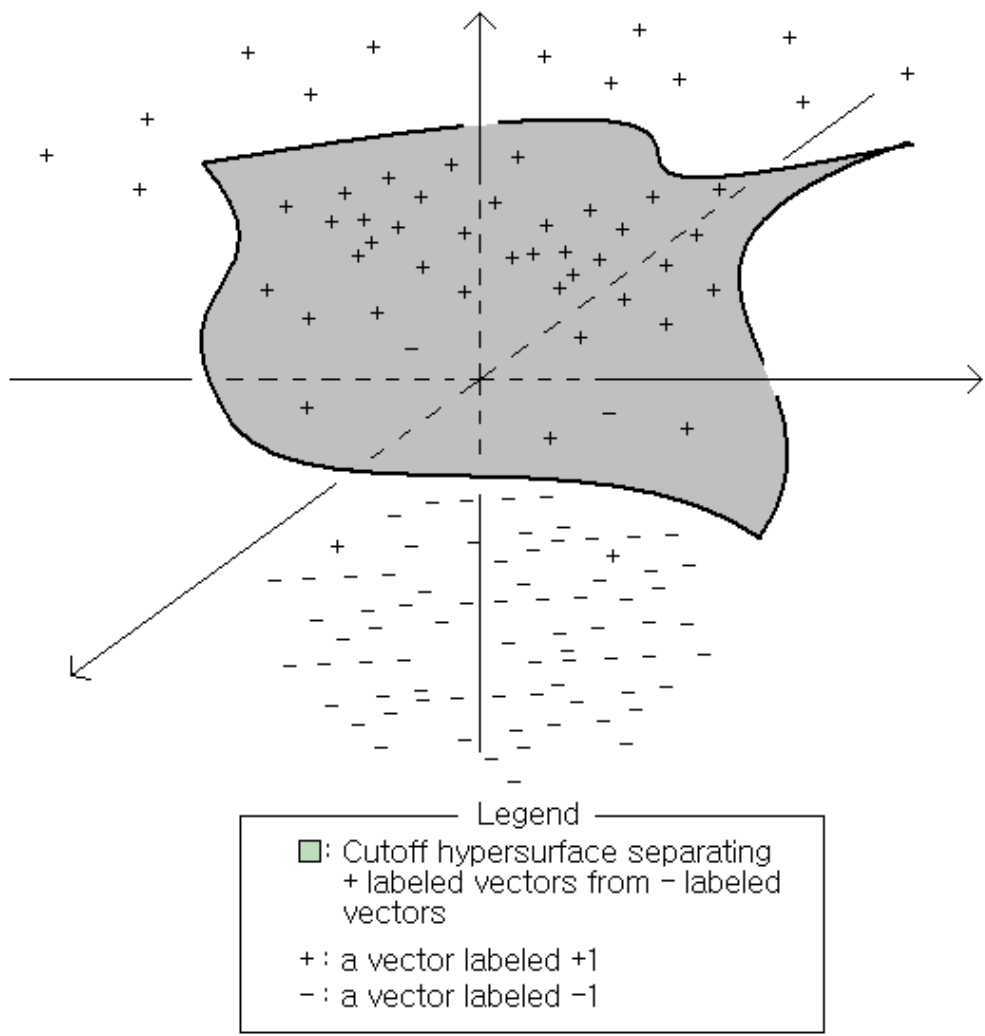

**Fig. 1**

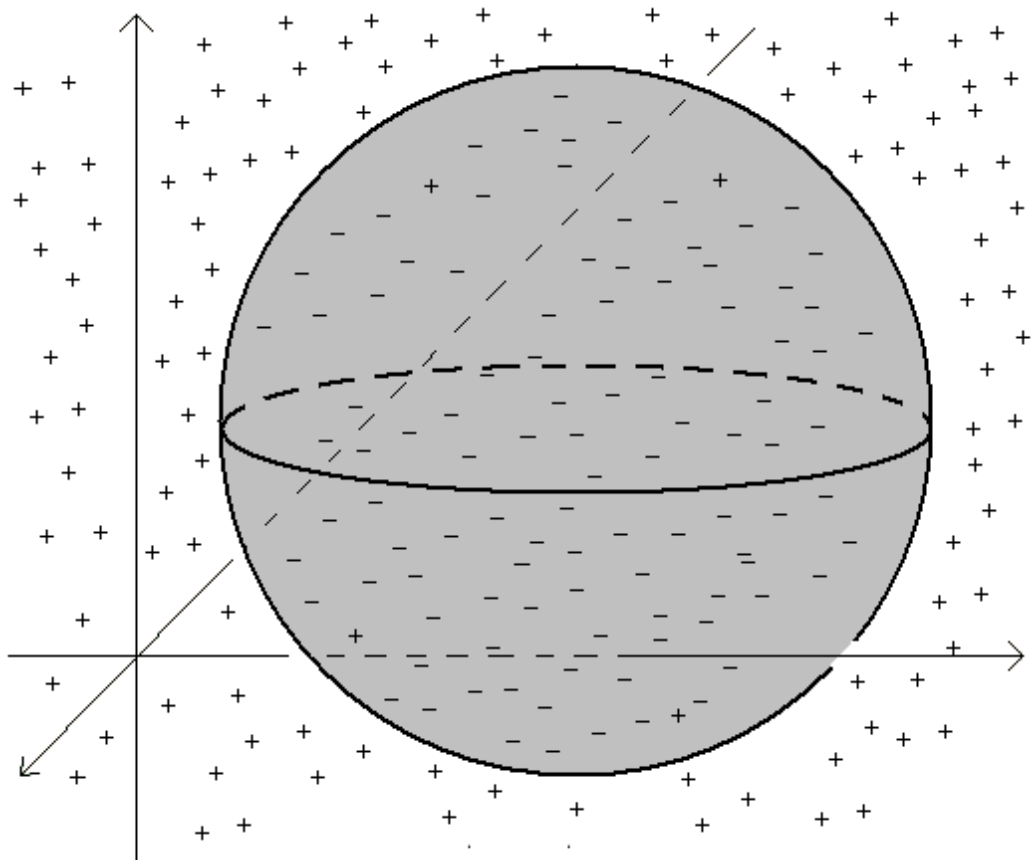

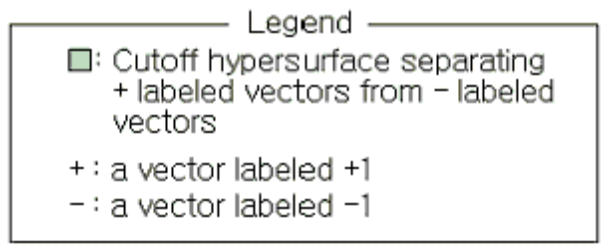

**Fig. 2**

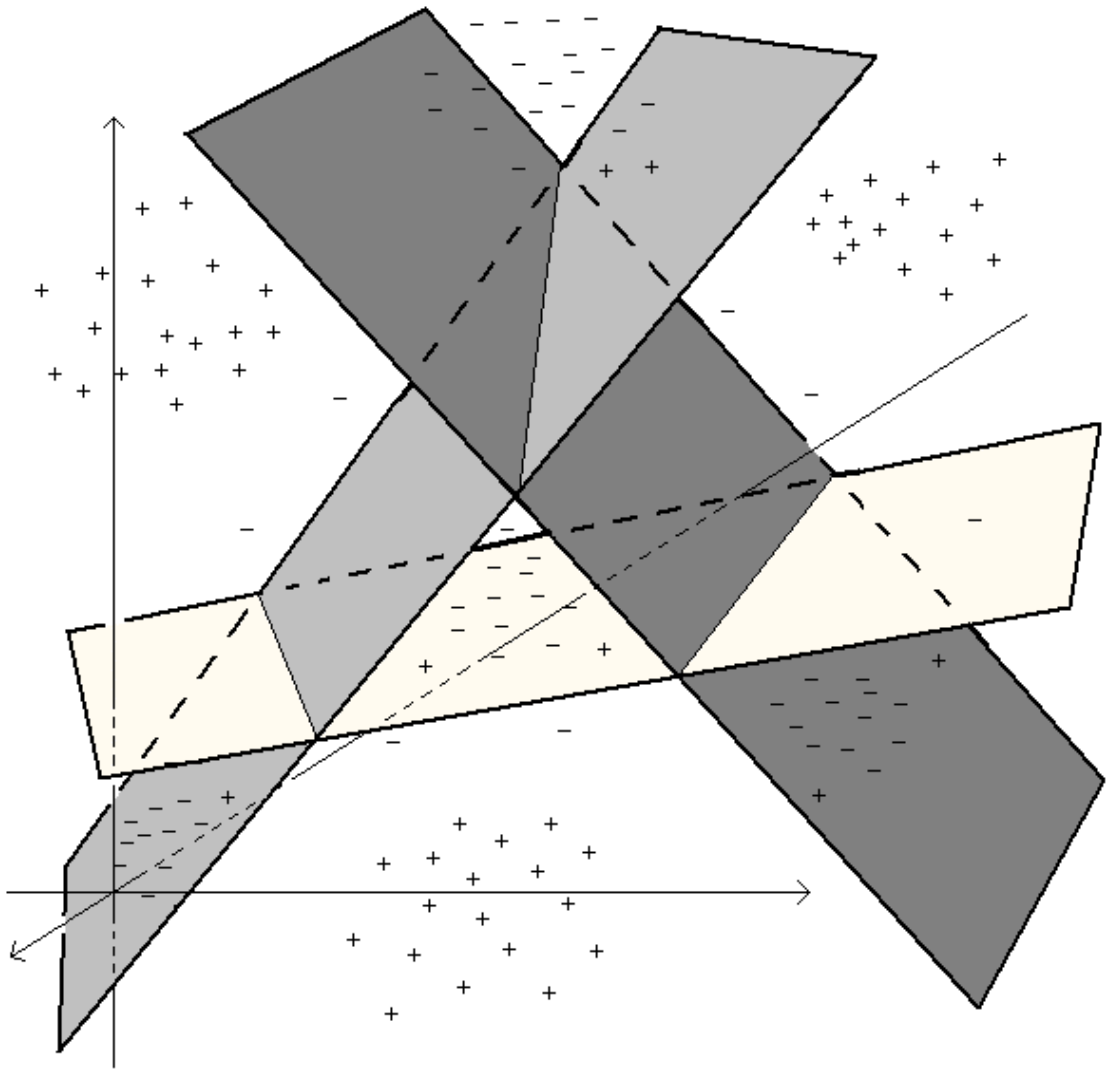

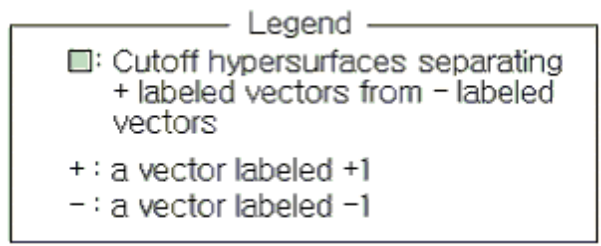

**Fig. 3**

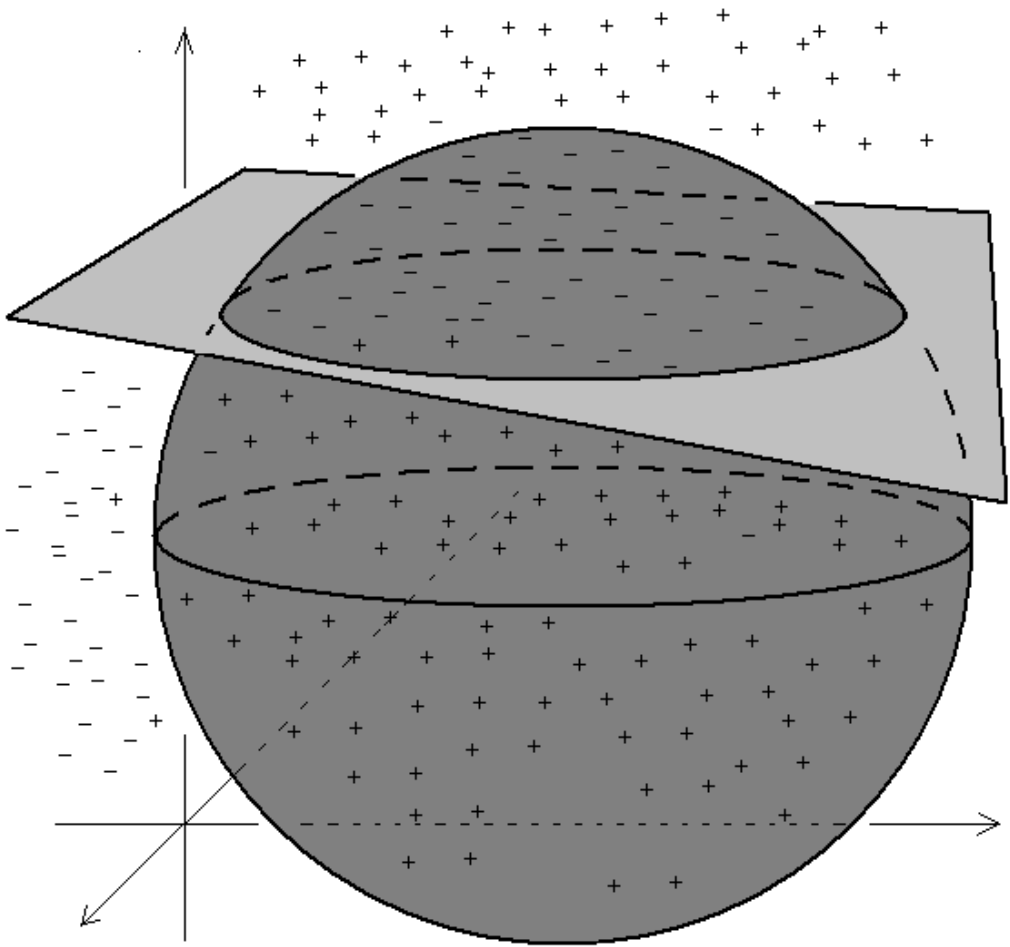

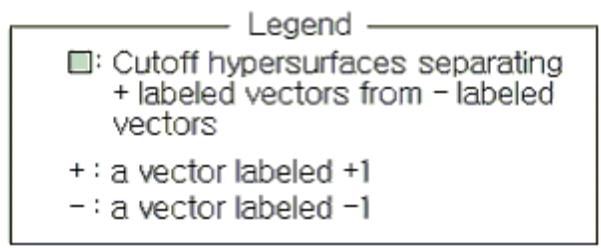

**Fig. 4**